\documentclass[useAMS,usenatbib]{mn2e}
\usepackage{epsfig,subfigure}

\title[IRAS04325+2402]{A multiwavelength view of the protostellar binary IRAS04325+2402: a case for turbulent
fragmentation}

\author[Scholz et al.]{A. Scholz$^{1,2}$\thanks{E-mail: aleks@cp.dias.ie}, K. Wood$^{2}$, D. Wilner$^{3}$,  
R. Jayawardhana$^{4}$, P. Delorme$^{2}$,
\newauthor{A. Caratti o Garatti$^{1}$, V. D. Ivanov$^{5}$, I. Saviane$^{5}$, B. Whitney$^{6}$}\\
$^{1}$ School of Cosmic Physics, Dublin Institute for Advanced Studies, 31 Fitzwilliam Place, Dublin 2, Ireland\\
$^{2}$ SUPA, School of Physics \& Astronomy, University of St. Andrews, North Haugh, St. Andrews, 
Fife KY16 9SS, United Kingdom\\
$^{3}$ Harvard-Smithsonian Center for Astrophysics, 60 Garden St., MS 42, Cambridge, MA 02138, USA\\
$^{4}$ Department of Astronomy \& Astrophysics, University of Toronto, 50 St. George Street Toronto, 
ON M5S 3H4, Canada\\
$^{5}$ European Southern Observatory, Ave. Alonso de Cordova 3107, Casilla 19, Santiago 19001, Chile\\
$^{6}$ Space Science Institute, 4750 Walnut St. Suite 205, Boulder, CO 80301, USA
}

\begin{document}

\date{Accepted. Received.}

\pagerange{\pageref{firstpage}--\pageref{lastpage}} \pubyear{2002}

\maketitle

\label{firstpage}

\begin{abstract}
IRAS04325+2402 (herafter IRAS04325) is a complex protostellar system hosting two young 
stellar objects (AB and C in the following) at a separation of 1250\,AU. Here we present 
new deep Gemini imaging and spectroscopy for the system covering the wavelength regime 
from 1-12$\,\mu$m as well as Sub-Millimeter Array interferometry at 870$\,\mu$m, in combination 
with Spitzer and literature data. Based on this rich dataset we provide a comprehensive picture of
IRAS04325 over scales from a few AU to several parsec. Object AB is a low-mass star with 
a disk/envelope system and an outflow cavity, which is prominently seen in infrared images. 
Object C, previously suspected to be a brown dwarf, is likely a very low mass 
star, with an effective temperature of $\sim 3400$\,K. It features an edge-on disk and an elongated 
envelope, and shows strong indications for accretion and ejection activity. Both objects 
are likely to drive parsec-scale molecular outflows. The two objects are embedded in an 
isolated, dense molecular cloud core. High extinction, lack of X-ray emission, and relatively 
high bolometric luminosity argue for a very young age below 1\,Myr. The disk/outflow systems 
of AB and C are misaligned by $\sim 60$\,deg against each other and by 80 and 40\,deg 
against the orbital plane of the binary. The system might be a good case for primordial 
misalignment, as opposed to misalignment caused by dynamical interactions, because the outflow
direction is constant and the realignment timescale is likely larger than the system age. This  
favours turbulent fragmentation, rather than rotational fragmentation, as the formation scenario. 
We show that the spectral energy distributions and images for the two objects can be reproduced 
with radiative transfer models for disk/envelope systems. Our analysis provides reassurance in 
the established paradigm for the structure and early evolution of YSOs, but stresses the importance 
of developing 3D models with sophisticated dust chemistry.
\end{abstract}


\section{Introduction}
\label{s1}

The study of protostars and their environment requires the analysis of radiation across a broad range of
wavelengths, including the infrared and the submm/mm regime. At the same time, it is desirable to achieve
high spatial resolution to characterize the small-scale structures close to the sources 
\citep[see review by][]{2007prpl.conf..523W}. However, only a few objects have been studied at uniform and 
high resolution over the full wavelength domain of relevance (1-1000$\,\mu$m). Testing the current 
paradigms for collapse, accretion/outflow, and disk evolution with high-resolution multi-wavelength 
studies of individual objects is thus of major importance.

\begin{figure}
\includegraphics[width=4.1cm,angle=0]{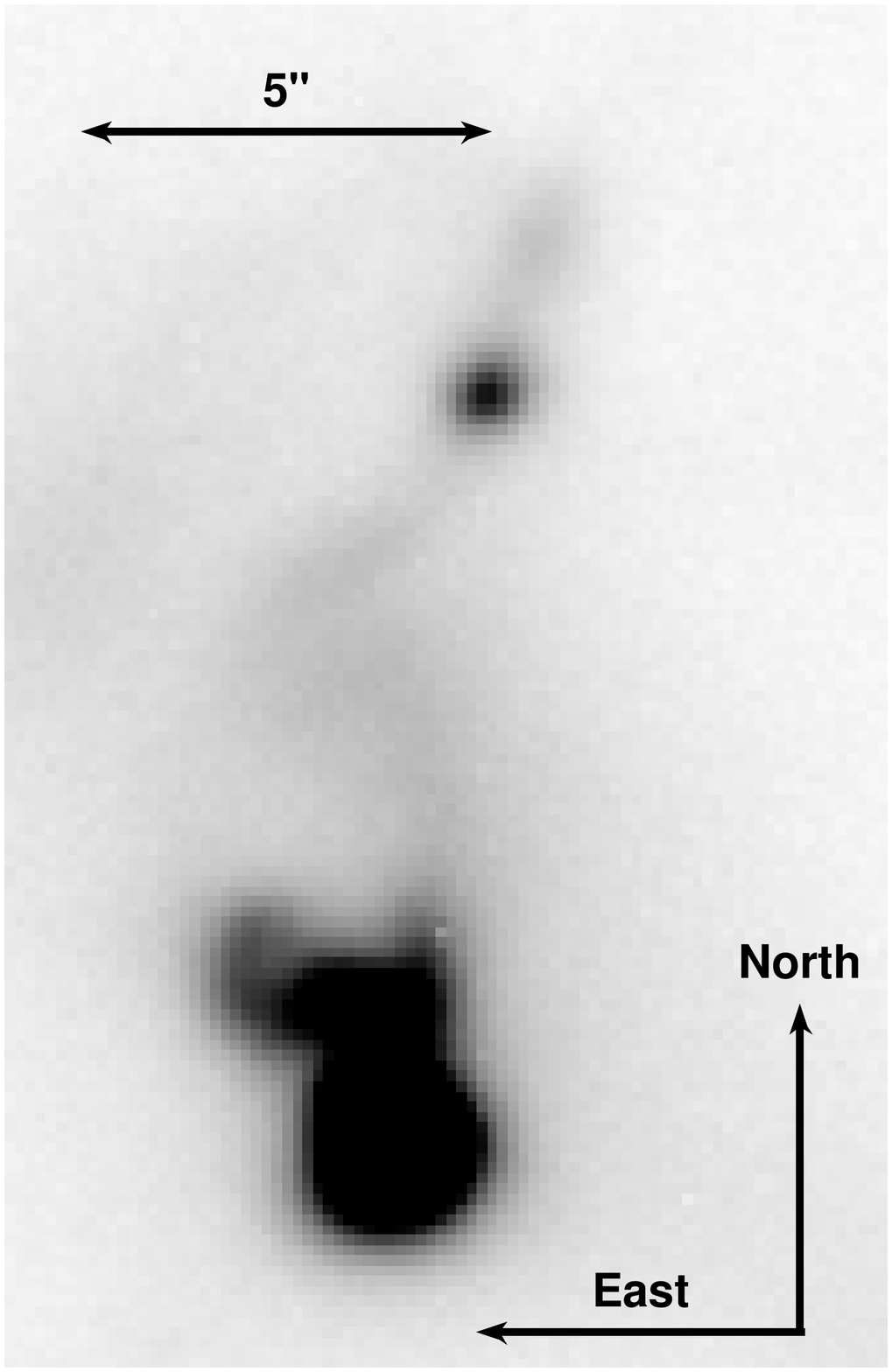} \hfill
\includegraphics[width=4.1cm,angle=0]{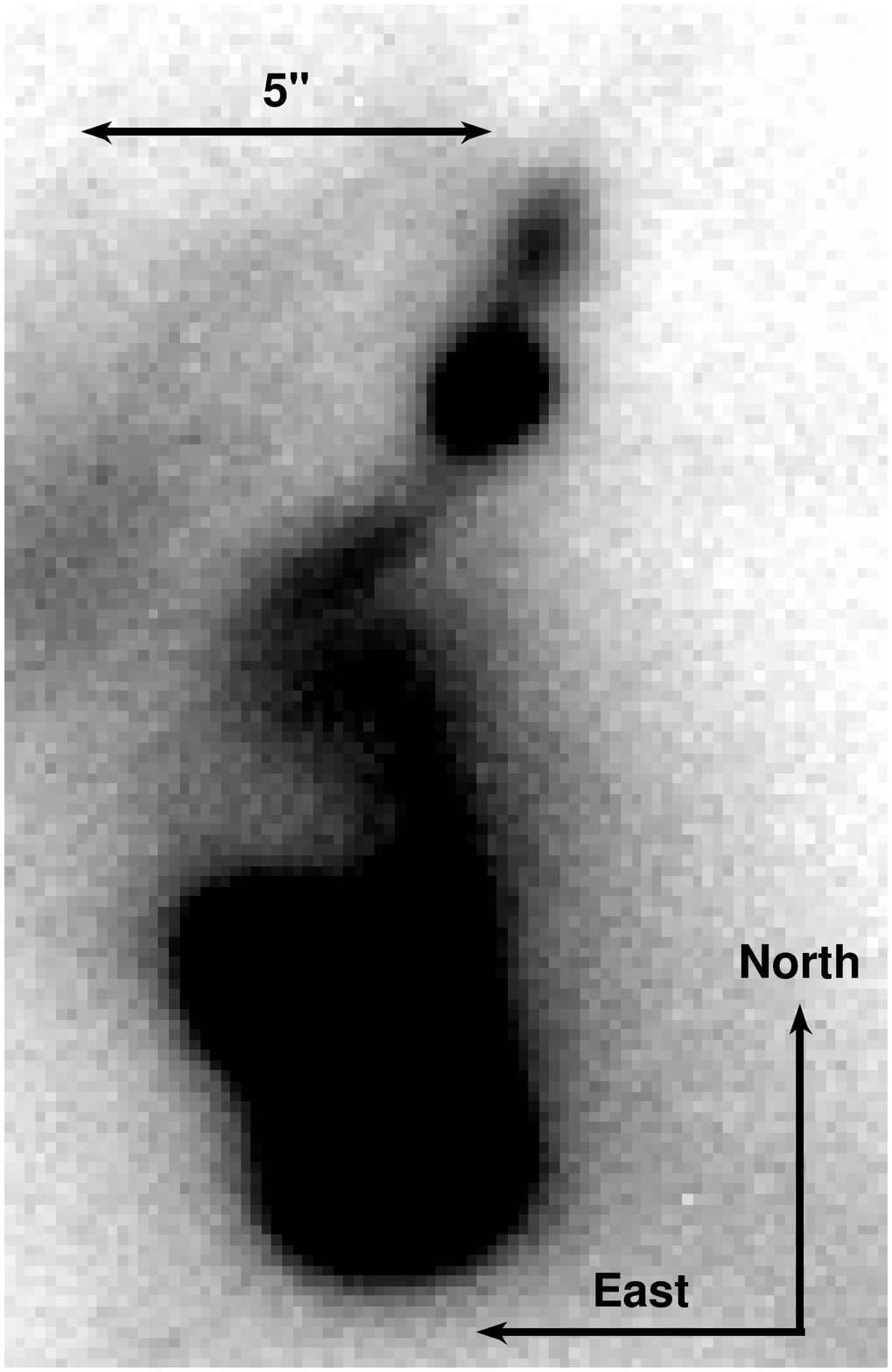} \\
\caption{K-band close-up view of IRAS04325 obtained with Gemini-North and NIRI. The image is shown in two 
different intensity scales: the left panel shows the point sources and bright extended emission close 
to the southern source AB, and the right panel illustrates the faint structure between AB and the northern 
source C.
\label{f0}}
\end{figure}

In this paper we present a case study for the protostellar system IRAS04325+2402 (hereafter: IRAS04325), shown
in Fig. \ref{f0}. Originally identified as LDN\,1535 in the catalogue of dark clouds by \citet{1962ApJS....7....1L}, 
it has been studied extensively over three decades (117 publications according to SIMBAD).
A number of publications starting in the 1980s has focused on the molecular gas in L1535 and the large-scale 
environment \citep[e.g.][]{1982A&A...111..339U,1984ApJ...283..140G}. After the identification of an IRAS source 
associated with this cloud \citep{1988MNRAS.235..139P,1991MNRAS.251...63P}, the object has been analysed in the 
infrared domain \citep[e.g.][]{1994ApJS...94..615H} and in the submm/mm continuum 
\citep[e.g.][]{1991ApJ...382..555L,1996ApJ...466..317O,1996A&A...309..827S}, tracing the dust in the system. 
IRAS04325 was found to harbour a protostar with subsolar luminosity and a disk seen at high inclination \citep{1993ApJ...414..676K,1993ApJ...414..773K}, embedded in a cloud core with a diameter of several tens 
of thousands of AU. In addition, large-scale outflow activity has been identified 
\citep[e.g.][]{1987ApJ...321..370H,1992ApJ...400..260M,2001AJ....121.1551W}. The source is 
established as a prototypical small-scale site of star formation.

Until 1999, the structures within the IRAS beam have not been resolved, which results in
various problems. The near-infrared images clearly show a bright nebulosity, indicating that the fluxes from IRAS and 
single-dish observations are contaminated by extended emission. As a result the properties of the central 
source and its close environment are poorly constrained. Confusingly, the direction of one of the large-scale 
outflows does not agree with the orientation of the infrared nebulosity, interpreted as outflow cavity, as pointed 
out by \citet{1993ApJ...414..676K}. \citet{1999AJ....118.1784H} presented HST imaging of the object, revealing complex 
substructure. In particular, the images show the presence of a second well-separated point source 
(IRAS04325C), which has a resolved disk seen edge-on. The main light source IRAS04325AB is speculated to be 
a binary. In addition there are various scattered light features.

Originally fueled by the suspicion the second source IRAS04325C might be a proto brown dwarf, we have conducted
our own observations of the system. In our first paper \citep{2008ApJ...681L..29S}, we detect the second component with
1.3\,mm interferometry and show the first infrared spectrum for this object. Here we present a comprehensive
dataset for the full IRAS04325 system, including new near- and mid-infrared imaging, near-infrared spectroscopy for 
both components, and submillimeter interferometry. We consistently assume a distance of 150\,pc for the Taurus
star forming region \citep{2007ApJ...671..546L} when converting from apparent to physical separations. All 
physical distances are given in projection against the plane of the sky and should be seen as lower limits 
to the actual distances.

\section{Observations and data reduction}
\label{s2}

\subsection{Near-infrared imaging}
\label{s21}

Near-infrared images in J-, H-, and K'-band were taken on the 23 August 2008 with the NIRI camera 
\citep{2003PASP..115.1388H} at the Gemini-North 8.1\,m telescope\footnote{The Gemini Observatory is operated 
by the Association of Universities for Research in Astronomy, Inc., under a cooperative agreement with the 
NSF on behalf of the Gemini partnership: the National Science Foundation (United States), the Science and 
Technology Facilities Council (United Kingdom), the National Research Council (Canada), CONICYT (Chile), 
the Australian Research Council (Australia), Ministério da Ciência e Tecnologia (Brazil) and Ministerio de 
Ciencia, Tecnología e Innovación Productiva  (Argentina)}. All Gemini observations for IRAS04325 were taken
in queue mode in the band-2 program GN-2008B-Q-73. We used the f/6 camera with a scale of 0\farcs12/pix 
and $120" \times 120"$ FOV. A 7-step dither pattern was carried out in each band with 15"
offsets and 10\,sec integration time per position. The total on-source time in each band was 70\,sec. 
The seeing measured from 1-3 well-detected point sources outside IRAS04325 was 0\farcs5. 

The reduction of the NIRI J-, H- and K'-band images proceeded as follows: a sky image for each filter was 
created using the dithered science images. The sky was then subtracted to each science image using the ESO 
eclipse package \citep{1997Msngr..87...19D}. We carried out a flatfield correction using domeflats. The 
alignment and coaddition of the images was done with the programs Scamp \citep{2006ASPC..351..112B} and 
Swarp \citep{2002ASPC..281..228B} to produce the final science stacks. 

In the same Gemini program, we obtained L'- (3.8$\,\mu$m) and M'-band (4.7$\,\mu$m) images with NIRI's f/32 camera 
(0\farcs02/pix) and a FOV of $22.4" \times 22.4"$. In the L-band a five step dither pattern with 3" offsets was 
executed twice with 20 coadded 1\,sec exposures at each position, which gives a total on-source time of 200\,sec. 
The same pattern was carried out 13 times in the M-band with 30 coadded 0.5\,sec exposures, in total 25\,min. These 
observations were obtained on the 3rd of December 2008, including a set of images for the bright A0 star HD36719 for 
calibration purposes. All data was taken at low airmass ($<1.2$); the airmass difference between science and 
calibration target was $\la 0.1$.

For the reduction of the L- and M-band data we followed the standard recipes based on the tools in the Gemini/NIRI
package within IRAF. Flats and sky images were created based on the dithered science images, discarding the first frame.
In the L-band all nine available images were used for the sky frame. The M-band sequence is long enough for the sky to 
change significantly. For each individual frame we used the nine frames taken closest in time to create the sky image.
After subtracting the sky the science frames were divided by the normalized flatfield. The resulting images were 
mostly flat with the exception of the first and last two frames in the M-band suffering from imperfect sky
subtraction. These images with uneven background were discarded. To coadd the individual images we used the standard 
{\tt combine} routine in IRAF. The images of the standard star for L- and M-band were handled in exactly the same 
way.

\subsection{Mid-infrared imaging}
\label{s22}

IRAS04325 was observed with Gemini-North and Michelle \citep{1994SPIE.2198..715B} in the N'-band with a central 
wavelength of 11.2$\,\mu$m on the 3rd of November 2008. To sample the sky properly the standard procedure including 
chopping and nodding was used. In short, the pointing was chopped at 4.2\,Hz between the target and an adjacent 
position 15" to the east by moving the secondary mirror. In addition the whole telescope was nodded in the same 
direction in larger time intervals. The chop throw was chosen to avoid the extended emission from IRAS04325 from 
the sky field. The exposure time per frame was 0.02\,sec; in total we integrated $2\times 590$\,sec on source. 
The airmass for the two science integrations was 1.07 and 1.02. In addition, the standard star HD20893 was 
observed at airmass 1.02.

For the standard reduction we used the Gemini/MIDIR package in IRAF and followed the 'cookbook' for Michelle provided by 
Gemini. This includes the tasks {\tt miregister} to prepare and reformat the data cubes and {\tt mistack} to average by 
coadding the signal from each nod position and dividing by the number of frames. A residual pattern was removed by fitting
each column with a constant and subtracting this fit. Source C is weakly visible only in one of the resulting stacked 
frames -- the one taken at lower airmss. The peak intensity for object C is about three times the noise 
in the sky background. The signal does not improve significantly when coadding the two images. Only the image with detection 
was used for the photometry. The standard star frames were reduced in the same fashion as the science images.

\subsection{Near-infrared spectroscopy}
\label{s23}

Object AB was observed with SofI \citep{1997Msngr..88...11M} in spectroscopy mode at the ESO/NTT in La Silla. 
We obtained this spectrum and the associated calibration files from the ESO archive. The low-resolution 
Red Grism was used with coverage from 1.5 to 2.5$\,\mu$m and a 0\farcs6 slit, which results in a 
resolution of $R \sim 1000$. Two pairs of nodded integrations (ABBA) were observed with a nodding step of 
20" and 225\,sec exposure time per position. The total on-source time was 900\,sec. A telluric 
standard star (HIP\,23719, spectral type G1V) was observed immediately after the science target.

We carried out a standard reduction, including subtraction of the nodded pairs (A-B) to remove
the sky and flatfield correction. The four spectra were extracted with a 1\farcs2 aperture, 
wavelength calibrated using Xenon arc spectra, and coadded. The telluric spectrum was treated 
in the same way. The combined spectrum for IRAS04325AB was then divided by the telluric spectrum 
and multiplied by the solar spectrum. The final science spectrum is shown in Fig. \ref{f10}. 

We obtained a spectrum for object C on Oct 14/15 2008 with Gemini-North and NIRI, using the f/6 camera, 
the 6-pix slit mask (slit width 0\farcs72) and the H- and K-band grisms. 
This gave a resolution of $R \sim 500$ over a wavelength range of 1.4-2.5$\,\mu$m. The 
slit was aligned in E-W direction. The total on-source time of 3000\,sec per band was split in 10 single 
integrations. To correct for the variable background emission, the object was moved 
along the slit in steps of 2" between the single exposures. For telluric correction we observed the early 
A stars HIP17791 and HIP23088 in the same observing nights.


The NIRI spectra were extracted and calibrated using our own IDL procedures. The frames were flat fielded 
using an internal flat taken immediately after the science frames. Consecutive exposures are subtracted
to remove the sky background. The 10 individual spectra are then extracted and median-combined. 
The same operation is performed for the telluric calibration stars. From the standard spectra we derived a 
telluric spectrum by dividing through a black body spectrum with a temperature of 10000\,K and interpolating 
over the hydrogen lines. The science spectrum was divided by the telluric spectrum. The wavelength 
calibration was carried out based on the bright OH lines which are visible in the (unreduced) science 
images. The final H- and K-band spectra are shown in Fig. \ref{f4}. 

\subsection{Archival images}
\label{s24}

To complement our dataset, we used archival images for the Taurus region. Our target and its environment 
is covered by the TAP Survey, carried out with the near-infrared camera WFCAM at the United Kingdom Infrared 
Telescope\footnote{The TAP survey is hosted by the Joint Astronomy Center. Image reduction was carried out by 
the Cambridge Astronomical Survey Unit for the WFCAM Science Archive. For more information, see 
{\tt http://www.jach.hawaii.edu/UKIRT/TAP/}.} \citep{2008MNRAS.387..954D}. The survey provides a database of 
near-infrared images in the K-band and the narrow-band filter centered on the 1-0 S(1) hydrogen emission 
feature at 2.122$\,\mu$m (in the following called: H$_2$ band), which we use to look for shocked emission 
caused by molecular outflows.

In addition, we downloaded the pipeline-reduced PBCD Spitzer images from the 3rd delivery (DR3) of the 
Taurus Spitzer Legacy project (PI: Deborah Padgett, program id \#3584) from the NASA/IPAC 
Infrared Science Archive. This includes the IRAC data in the four channels at 3.6, 4.5, 5.8 and 
8.0$\,\mu$m, as well as the MIPS images at 24 and 70$\,\mu$m.

\subsection{Submillimeter interferometry}
\label{s25}

\begin{figure}
\includegraphics[width=8.0cm,angle=-90]{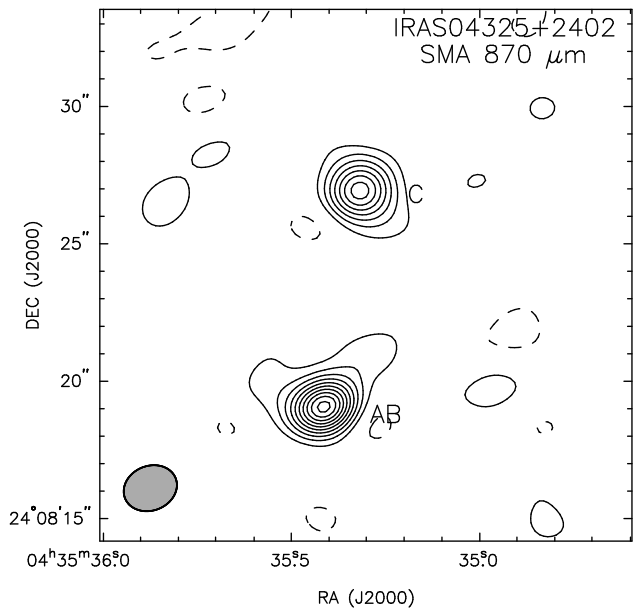} \hfill
\caption{SMA 870$\,\mu$m dust continuum image showing the two sources AB and C and associated 
extended emission. The ellipse in the lower left corner shows the $2\farcs0 \times 1\farcs6$ PA 
$-68$\,deg beam. The lowest contour and step are 4.4\,mJy/beam (twice the rms noise level). Negative
contours are dashed. \label{f9}}
\end{figure}

\begin{figure*}
\includegraphics[width=5.5cm,angle=-90]{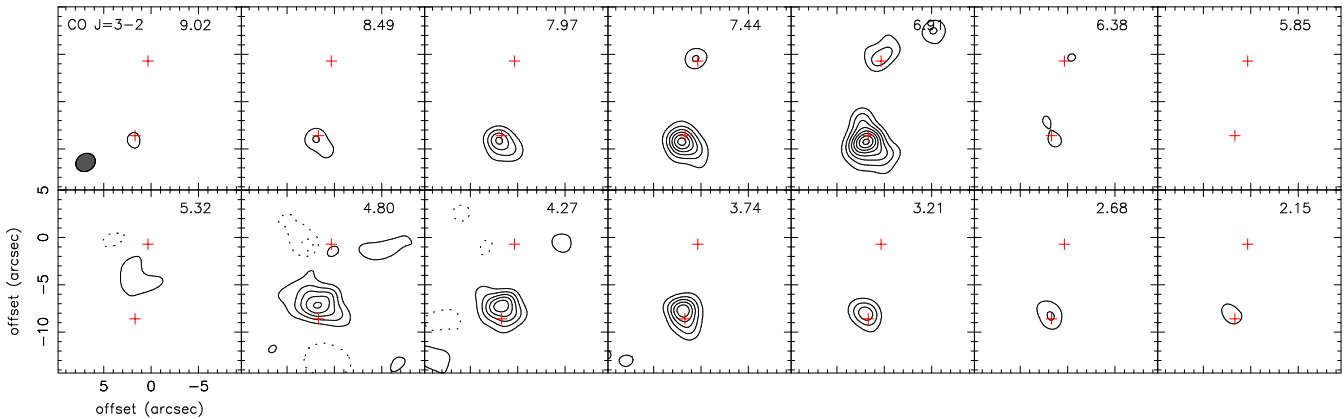} \hfill
\caption{Contour maps of the continuum-subtracted CO line observations with SMA. The contours are 
in steps of 1.5\,Jy/beam, negative contours are dashed. The LSR velocity in kms$^{-1}$ is indicated 
for each panel. The positions of the compact sources AB and C, as measured in the continuum map, are 
marked with crosses. \label{f5}}
\end{figure*}

A field centered on IRAS04325 was observed with the 
SMA\footnote{The Submillimeter Array is a joint project between the Smithsonian Astrophysical Observatory and 
the Academia Sinica Institute of Astronomy and Astrophysics and is funded by the Smithsonian Institution and 
the Academia Sinica.} on 20 October 2008 using eight antennas in a compact configuration that provided baseline 
lengths of 11 to 77\,m. The weather was good, with system temperatures of 200 to 250 K (DSB) near transit. 
The phase center was RA $04^{h}35^{m}35\fs29$, DEC $+24^{o}08'27\farcs6$ (J2000), which is the position of
IRAS04325C. The correlator was configured for the full 2\,GHz bandwidth, including one spectral chunk of 
82\,MHz in the middle of the upper sideband devoted to the CO J=3-2 line at 345.786\,GHz with channel spacing 
0.20\,MHz (0.176 km~s$^{-1}$). The SMA primary beam FHWM is $35''$ at this frequency. Calibration of complex 
gains was performed by interleaving 3 minute observations of the quasars 3C111 (4.30\,Jy) and J0530+135 (0.84\,Jy) 
with 15 minute observations of IRAS04325 over the hour angle range $-5$ to $+5$. The flux scale was set using 
observations of Uranus, with an estimated uncertainty better than 20\%. The passband was calibrated using 
observations of the strong sources 3C454.3 and Uranus. The data calibration was done with the MIR/IDL software, 
followed by standard imaging and deconvolution tasks within the Miriad package. For the continuum image, the 
beam size (robust=0) was $2\farcs0 \times 1\farcs6$ at PA $-68$\,deg, and the rms noise 2.2\,mJy/beam, after
combining the lower and upper sidebands (effective bandwidth $\sim4$~GHz). For the CO line images, the beam 
size (natural weight) was $2\farcs1 \times 1\farcs8$ at PA $-56$\,deg, with rms noise 0.26\,Jy/beam in each
spectral channel. The continuum image is shown in Fig. \ref{f9}; the CO line maps for a sequence of 
velocity bins are plotted in Fig. \ref{f5}. Since the interferometer observations are sensitive only to small 
scale structure ($\la 10''$), the sources seen in the CO maps are not isolated and should be seen as local 
overdensities and temperature enhancements.

\section{Quantitative analysis}
\label{s3}

\subsection{Object AB: NIR spectroscopy}
\label{s31}

\begin{figure}
\includegraphics[width=6.1cm,angle=-90]{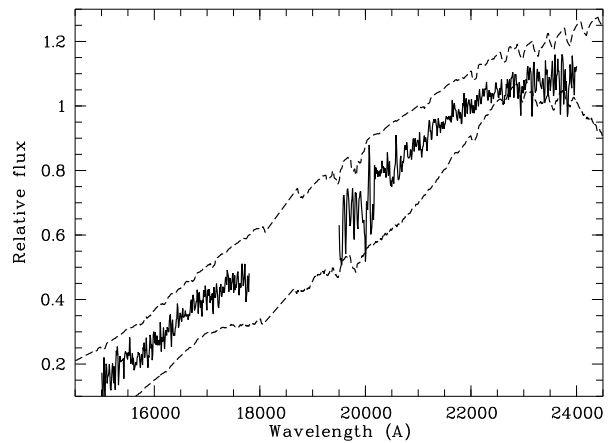} \hfill
\caption{Spectrum for AB obtained with NTT/SOFI. Overplotted with dashed lines are the spectra for a 
K7V star (above) and a M4V star (below the science spectrum), from the IRTF/SpeX Spectral Library 
\citep{2009arXiv0909.0818R}. Both comparison spectra are reddenened with $A_V = 30$\,mag.
\label{f10}}
\end{figure}

The spectrum of the source AB is red and entirely featureless (see Fig. \ref{f10}), without the broad water 
absorption bands usually seen in late-type objects \citep{2005ApJ...623.1115C}. The slope from 1.5 to 
2.3$\,\mu$m is well approximated by a straight line. There may be a hint of CO absorption bands at $>2.3\,\mu$m. 

The lack of features precludes a detailed spectral analysis. Instead we simply compare the slope with reddened
spectra for late-type objects from the IRTF/SpeX Spectral Library \citep{2009arXiv0909.0818R}. G-, K- and
early M-type spectra reddenend with $A_V \sim 30$ provide a good match. As an example we show in Fig. \ref{f10} 
the spectrum for the K7V star HD237903. Towards mid-M spectral types the fit becomes progressively worse, no matter
which extinction is chosen. In particular, the slope over H- and K-band is no longer constant, due to the water 
absorption feature between the two bands. In addition, the CO bands and water absorption cause the flux level to 
drop at $>2.2\,\mu$m. The mismatch is shown in Fig. \ref{f10} where we compare with the spectrum of the M4V 
star Gl\,299 reddened with $A_V=30$\,mag. Thus, a spectral type later than early M seems unlikely.

This limit on the spectral type, however, needs confirmation, particularly because the spectrum 
may be affected by additional continuum emission from the disk or accretion. The slit orientation in E-W direction 
avoids the scattered light features north and south of the source (see Fig. \ref{f2}), but veiling from the inner
disk (within 0\farcs5) cannot be excluded. Lacking high-resolution spectroscopy, we cannot put limits on
the veiling. Typical excess continuum in T Tauri stars makes the near-infrared spectrum 
redder \citep{1999A&A...352..517F}, i.e. the spectral type may be earlier than estimated or the 
extinction may be overestimated.

\subsection{Object C: NIR spectroscopy}
\label{s32}

\begin{figure*}
\includegraphics[width=6.1cm,angle=-90]{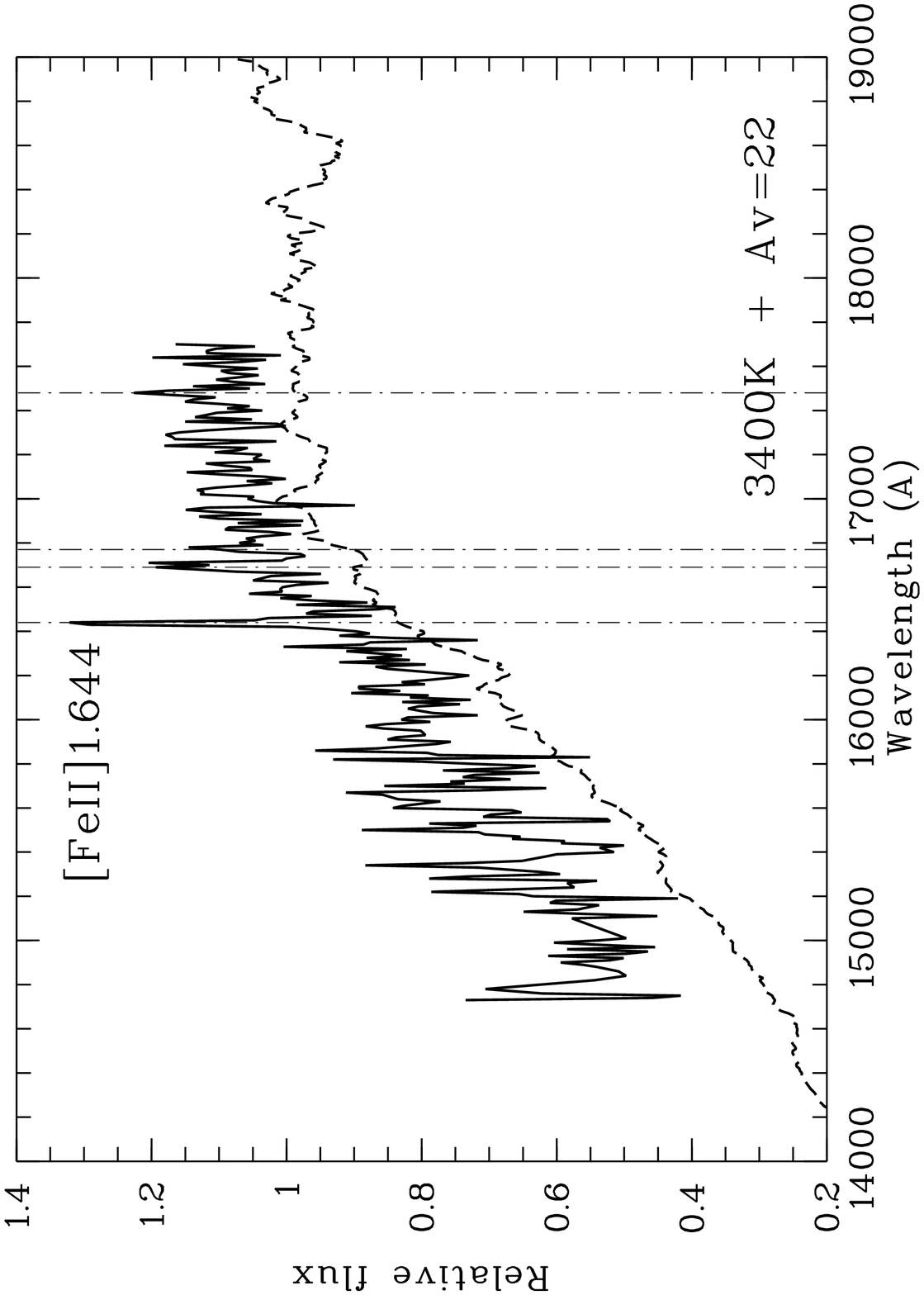} \hfill
\includegraphics[width=6.1cm,angle=-90]{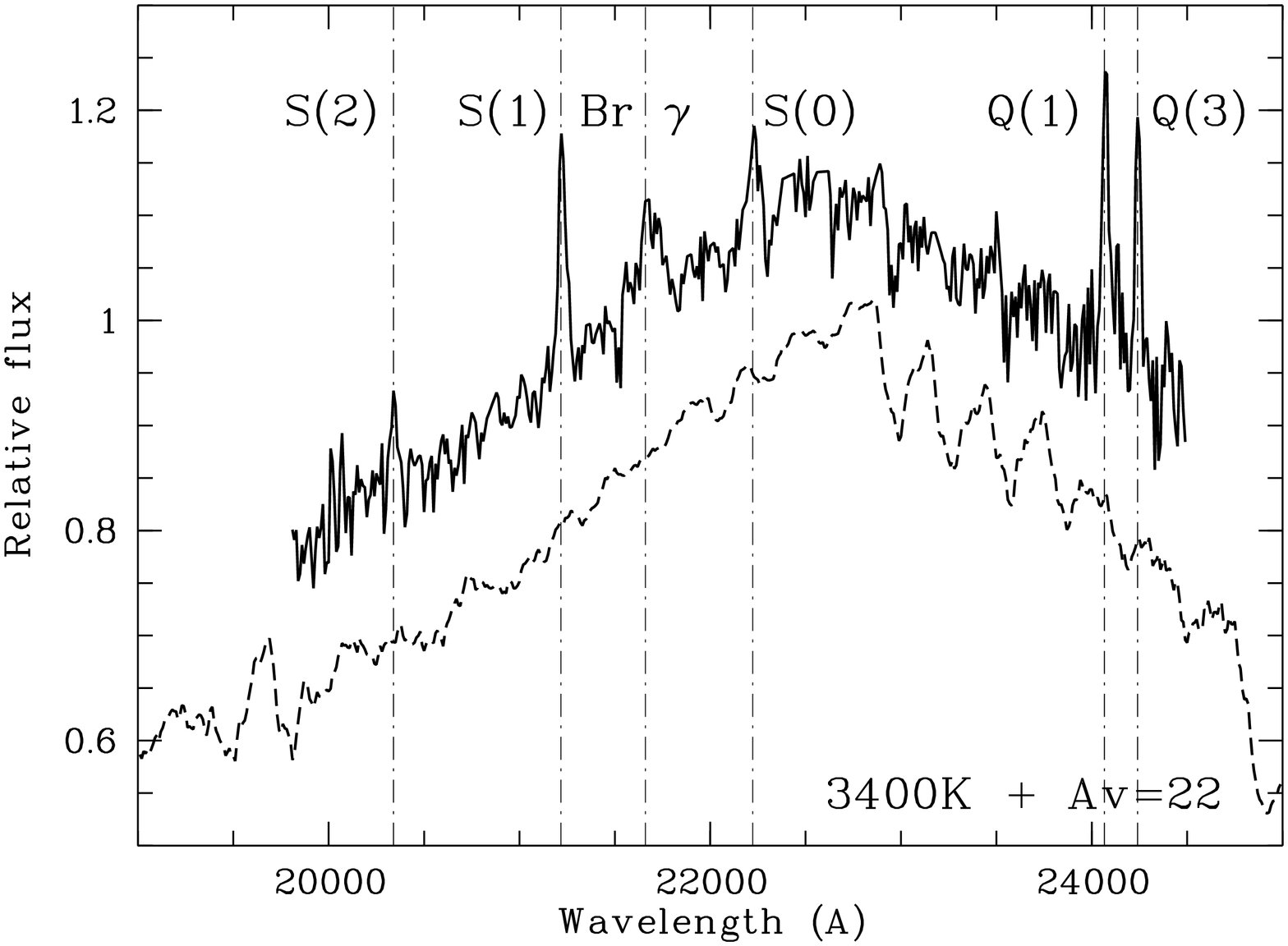} 
\caption{H- and K-band spectrum from Gemini/NIRI (solid line) in comparison with a model (DUSTY, 3400\,K, $\log g = 4.0$, 
$A_V=22$\,mag, see \citet{2001ApJ...556..357A}). The model (dashed line) is shifted in y-direction for clarity. The 
emission lines detected on the 3$\sigma$ level are marked with dashdotted lines (see text for more information). 
\label{f4}}
\end{figure*}

The spectrum of object C shows features of the photosphere, in particular a flattening at the red 
end of the H-band and a triangular shape in the K-band (Fig.\ref{f4}). The spectra in H- and K-band were 
compared separately to model spectra. We neglect that the spectra might be affected by scattering and 
veiling. Both effects are expected to be smoothly and slowly varying functions of wavelength, thus they 
should not affect the spectrum in individual bands significantly.

For the comparison we used a series of AMES-DUSTY model spectra \citep{2001ApJ...556..357A} with effective 
temperatures ranging from 2000 to 4000\,K and $\log g = 3.5$ or $4.0$, as expected for young 
stars. We varied $T_\mathrm{eff}$ and $A_V$ and aimed to obtain a consistent solution for the two bands.
We find a decent match between observed and model spectra for $T_\mathrm{eff} = 3400 \pm 200$\,K and 
extinction $A_V = 22\pm 2$\,mag. As seen in Fig. \ref{f4} this fits H and K bands equally well. 

It is impossible to match the observed spectrum with a model for an object close or below the substellar
boundary ($T_\mathrm{eff} \la 3000$\,K), as claimed by \citet{2008ApJ...681L..29S}, particularly not in the 
H-band. Young brown dwarfs exhibit a characteristic triangular shaped H-band spectrum due to water absorption
\citep{2006ApJ...639.1120K,2006ApJ...653L..61B,2009ApJ...702..805S} which is not seen in our data. We conclude 
that IRAS04325C is unlikely to be a brown dwarf; its spectrum is more consistent with being a very low-mass 
star with an effective temperature around 3400\,K or spectral type of M3-M4 \citep{2008ApJ...689.1127M}.

Various emission features associated with molecular hydrogen are clearly detected in the K-band, 
particulary 1-0 S(2) at 2.034, 1-0 S(1) at 2.122, 1-0 S(0) at 2.223, 1-0 Q(1) at 2.407, and 1-0 
Q(3) at 2.424$\,\mu$m. In addition, there is a Br $\gamma$ emission feature at 2.166$\,\mu$m, 
commonly associated with ongoing accretion \citep[e.g.][]{2006A&A...452..245N}. The H-band shows 
the [FeII] lines at 1.644, 1.669, 1.677, and 1.748$\,\mu$m (the latter might be blended with a 
hydrogen feature). While the lines of molecular hydrogen can originate in the disk or in a jet, the 
[FeII] emission can only occur in the jet. Thus, our spectra prove ongoing accretion and outflow 
activity in IRAS04325C. From the way we have analysed the spectra it is certain that the emission 
originates in an area within the PSF of the point source, i.e. within a radius of $0\farcs5$ (75\,AU) 
around the source.

The [FeII] lines provide a constraint on the electron density $n_e$ in the jet 
\citep[e.g.][]{2002A&A...393.1035N,2008A&A...487.1019G}. The flux ratio between the well-detected 
lines at 1.644 and 1.677$\,\mu$m is 2.9, which gives $n_e > 10^5$\,cm$^{-3}$. Note that the 
[FeII] line ratio saturates for larger densities. Consistently, the lower limits to the flux ratios of 
1.544 vs. 1.644$\,\mu$m and 1.600 vs. 1.644\,$\mu$m yield densities $>0.5 \cdot 10^5$ and $>10^4$\,cm$^{-3}$, 
respectively. Such high values are usually found at the base of the jet, while the density typically 
drops further away from the source \citep[e.g.][]{2006ApJ...641..357T,2008A&A...487.1019G}. Thus the 
analysis confirms that the [FeII] emission originates in a jet close to IRAS04325C.

The Br $\gamma$ line is used to derive an estimate for the mass accretion rate following 
\citet{2006A&A...452..245N}. The approximate equivalent width in this line is $7 \pm 2$\,\AA. 
Scaling with the K-band magnitude of 8.5\,mag for a 3400\,K star at 1\,Myr from 
\citet{1998A&A...337..403B} gives a line luminosity of $\log (L_{\mathrm{Br\gamma}} / L_{\odot}) \sim -4.1$.  
With the empirical relations by \citet{2004AJ....128.1294C} this translates into an 
accretion luminosity of $\log (L_\mathrm{acc} / L_{\odot}) \sim -0.8$ 
\citep[see also][]{1998AJ....116.2965M}. The mass accretion rate is then 
$\dot{M} = L_{\mathrm{acc}} R / G M$. Assuming a mass of 0.4\,M$_{\odot}$ and a radius 
of 1.5\,R$_{\odot}$ results in an accretion rate in the range of 
$(1.0 \pm 0.3) \cdot 10^{-8}\,$M$_{\odot}$\,yr$^{-1}$, comparable to accretion rates derived 
for T Tauri stars but higher than those for young brown dwarfs \citep{2005ApJ...626..498M}.

\subsection{Photometry of point sources}
\label{s33}

We derived magnitudes/fluxes for the two components of IRAS04325 from the Gemini near- and mid-infrared
images as well as from the submm continuum map.

For the JHK images we adopted an aperture of 1\farcs2 (10 pixel) and a background annulus between 1\farcs2 
and 1\farcs8 (10-15 pixel). The field contains one 2MASS source outside IRAS04325 with $J=17.98$, $H=16.41$, 
and $K=14.64$ which was used to shift the instrumental magnitudes into the 2MASS system. The resulting 
values are listed in Table \ref{phot}. In the J-band there is diffuse emission at the position of object 
C, but no clear point source; we consider the flux measurement to be an upper limit. The previously published 
photometry of IRAS04325 has been obtained with larger apertures \citep{1999AJ....118.1784H,2008AJ....135.2496C}. 
Hence, these literature fluxes are significantly larger than ours because they include extended emission. 

For the L'-, M'-, and N'-band photometry we used apertures of 1\farcs0 or 1\farcs2 and a sky annulus similar
to the one for JHK. In these bands the extended emission is less problematic, thus the choice of the sky annulus 
does not significantly affect the results. The L'/M'-band magnitudes were shifted to the standard system based 
on the values for the calibration star HD36719, which has 5.4\,mag throughout this wavelength regime 
\citep{2003MNRAS.345..144L}. For the N'-band magnitudes we used the star HD20893 for calibration, for which 
an N'-band flux of 4.3\,Jy is reported on the Gemini website\footnote{{\tt http://www.gemini.edu/sciops/instruments/ 
mir/MIRStdFluxes.txt}}, 
based on \citet{1999AJ....117.1864C}.

The fluxes in the 870$\,\mu$m continuum map, measured within a 3\farcs0 aperture, are 55 and 
41\,mJy, for AB and C respectively.

The uncertainties were estimated by varying the size of the aperture and the sky annulus and
checking the changes in the fluxes. Combined with the uncertainty in the fluxes used for the calibration,
this mostly yields errors $\la 10$\%. The exceptions are the L'- and M'-band images for object C, where the
errors are substantially larger.

\begin{table}
\caption{IR photometry for the point sources in IRAS04325. If not specified, the uncertainty in 
the fluxes is $\sim 10$\%; this includes the calibration error. For more details see Sect. \ref{s33}.
The fluxes at 1300\,$\mu$m are from \citet{2008ApJ...681L..29S}.}
\label{phot}
\begin{tabular}{ccrrcc} 
\hline
Band     & $\lambda$ ($\mu$m) & AB          &  C            & unit    & aperture (") \\
\hline
J        & 1.25                & 16.54        &  $<18.6$      & mag     & 1.2   \\
H        & 1.65                & 14.37        &  16.92        & mag     & 1.2   \\
K'       & 2.15                & 11.87        &  14.75        & mag     & 1.2   \\
L'       & 3.78                & 10.36        &  $13.4\pm0.2$ & mag     & 1.0   \\
M'       & 4.68                & 9.37         &  $12.3\pm0.4$ & mag     & 1.0   \\
N'       & 11.2                & 65           &  $3.4\pm1.1$  & mJy     & 1.2   \\
         & 870                 & 55           &  41           & mJy     & 3.0   \\
	 & 1300                & $10\pm ^3_1$ &  $10\pm ^3_1$ & mJy     & 2.0   \\
\hline			     
\end{tabular}			     
\end{table}	


\begin{figure}
\includegraphics[width=6.0cm,angle=-90]{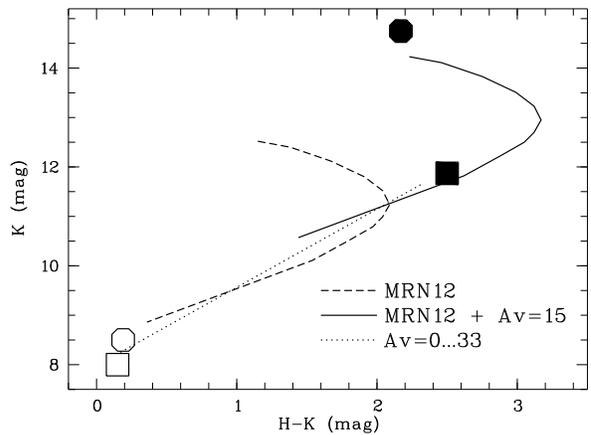} \hfill
\caption{Colour-magnitude diagram illustrating the interpretation of the near-infrared photometry. Filled symbols
show measured photometry; empty symbols the approximate photospheric fluxes obtained by comparing the
constraints on $T_{\mathrm{eff}}$ with the 1\,Myr evolutionary track by \citet{1998A&A...337..403B}. Squares
correspond to object AB and octagons to object C. The dotted line is the reddening path for $A_V=1-33$ with 
the \citet{1990ARA&A..28...37M} extinction law. The dashed line show the standard scattering model 
MRN12 from \citet{1993ApJ...414..773K} with inclinations ranging from $\cos i = 0.975-0.0$. The solid 
line is the scattering model plus extinction.
\label{f1}}
\end{figure}

In Fig. \ref{f1} we analyse the NIR colours and magnitudes of the two sources (filled symbols). In addition,
we estimated the photospheric colours and magnitudes by comparing the constraints on the effective
temperatures with the 1\,Myr evolutionary track by \citet{1998A&A...337..403B} (empty symbols). For 
source C, we used $T_{\mathrm{eff}} = 3400$\,K, as derived by comparing the spectrum to models 
(Sect. \ref{s32}). For object AB we only have a limit on the spectral type ($\la$ early M), which 
corresponds to $T_{\mathrm{eff}} \ga 3600$\,K \citep{2008ApJ...689.1127M}. Thus, the plotted flux
constitutes a lower limit as well. These photospheric datapoints may be uncertain in the y-axis 
by up to $\pm 1$\,mag due to the unknown age and the uncertainty in $T_{\mathrm{eff}}$, but this 
does not affect the following interpretation significantly. 

For object AB (square symbols) the colours are well explained by the extinction path (dotted line), 
indicating a line of sight extinction of $A_V \sim 30$\,mag, in agreement with our constraint from the 
spectrum (Sect. \ref{s31}). Thus, both spectroscopy and photometry are consistent with an extincted 
photosphere of a single low-mass star. For object C (octagons), however, it is clear that extinction 
alone cannot explain the $H-K$ colour. Object C appears to be 6\,mag fainter in the K-band image than 
expected from the photospheric fluxes derived from the spectrum. Scattered light models provide a much 
better fit. The solid line shows the model MRN12 from \citet{1993ApJ...414..773K} plus an extinction 
of $A_V = 15$\,mag, which allows to reproduce the K-band flux and the colour simultaneously. This 
analysis confirms that the object is mostly seen in scattered light through an edge-on disk.

In summary, the information from photometry and spectroscopy indicates that AB and C are both
low-mass stars. Since the mass of AB is likely to be higher than the mass of C and AB is located 
closer to the core of the nebula, AB is probably the most relevant center of infall in the 
IRAS04325 system.

\subsection{PSF fitting}
\label{s34}

It has been speculated that AB might be a binary (hence the name). Based on the HST images, 
\citet{1999AJ....118.1784H} argue that 'there is a double structure, possibly indicating two sources or 
possibly indicating a dark absorption lane running roughly east-west across the object.' In the HST images 
the separation between the two components is in the range of 0\farcs2 roughly in north-south direction, 
corresponding to $\sim 30$\,AU. The 'dark lane' would indicate an absorption feature in front of the 
object, possibly a disk with diameter of a few tens AU. 

Based on our K-band image, we test for the presence of a second source in AB by constructing a model PSF 
from the three well-detected field stars outside IRAS04325. All sources in the image are fit with
this model PSF, using {\tt daophot} within IRAF. The $\chi^2$ of the fit is $<2$ for all field stars, 
3.2 for object C, and 12.0 for object AB. The contour plot of the residuals (Fig. \ref{f2}) after
subtracting one PSF does not show any evidence for a second point source, and fitting two PSFs instead
of one does not improve the fit. The high $\chi^2$ is most likely caused by the strongly uneven 
background and not by a stellar companion. 

Thus, we prefer to interpret the double structure seen in the HST image as an indication for a disk 
seen at high inclination that bisects the image of the star, rather than the presence of a resolved 
companion. This finding is supported by the combined information from photometry 
and spectroscopy; for a binary we would expect a mismatch (i.e. a later spectral type than expected 
from photometry). The orientation of the 'lane' is roughly perpendicular to the outflow emanating 
from AB (Sect. \ref{s4}).

In the submm continuum image both AB and C are spatially resolved. The spatial structure is studied 
after discarding the data from baselines shorter than 44\,m, which effectively removes the structures larger 
than a few arcsec. We used the task {\tt uvfit} in the Miriad package to fit the visibilities of the 
sources. Various combinations of point sources and gaussian sources were tried for both objects. 

The northern source C is well matched by an elliptical gaussian with axes of 1\farcs2 and 0\farcs2 and a 
position angle of $\sim 50$\,deg (Fig. \ref{f9}). Note that the position angle of the beam is SE to NW, 
i.e. perpendicular to the orientation of the source. The parameters of the gaussian have large uncertainties,
but it is safe to conclude that the position angle is consistent with the orientation of the disk, as 
inferred from the HST images \citep[PA 30\,deg,][]{1999AJ....118.1784H}. The same HST images also 
constrain the radius of the disk to 0\farcs2 ($\sim 30$\,AU), which is not resolved by the SMA. 
The elongated structure seen in the submm data might be caused by a cold outer disk or by an elongated 
circumstellar envelope with a diameter of $\sim 1-2"$, i.e. 150-300\,AU. This size matches well with 
the constraint from the SED modeling (Sect. \ref{s52}). 

The southern AB source contains a compact component that is well matched by a point source, likely 
caused by a small-scale disk. In addition there is a contribution of spatially extended emission, 
mostly oriented in E-W direction over an area of 3-5".

\subsection{Astrometry}
\label{s35}

Based on the PSF fit discussed in Sect. \ref{s34} the relative positions of AB and C were determined with an accuracy
in the range of 0\farcs1 (1 pixel). The distance between the two sources measured from the peak position of the model PSF 
is 8\farcs32. Simply measuring the positions of the emission peaks also gives a consistent 
distance of 8\farcs3. From the coordinates given by \citet{1999AJ....118.1784H}, based on images from 
November 1997, we infer a distance between AB and C of 8\farcs23. Within the errorbar both measurements are 
consistent.\footnote{For completeness, the relative distance from the astrometry given by \citet{2008AJ....135.2496C} 
based on imaging from 2003-5 is 8\farcs03; the accuracy of their astrometry is not clear.}  The position angles from 
the available near-infrared images agree well: We measure 351.8\,deg (E of N), while the literature values are 
351.4 \citep{2008AJ....135.2496C} and 351.6\,deg \citep{1999AJ....118.1784H}, all with uncertainties of $\la 1$\,deg. 

These measurements constrain the relative proper motion between AB and C 
to $<9$\,mas\,yr$^{-1}$. For comparison, the average proper 
motion of young stars within 5\,deg of IRAS04325 is 23\,mas\,yr$^{-1}$ \citep{2005A&A...438..769D,2009ApJ...703..399L}, 
with a standard deviation of 10\,mas\,yr$^{-1}$. While the constraints on the proper motion for AB and C do not prove 
that the object is a physically bound system, they do confirm common membership in the Taurus association.

The angular distance between AB and C correspond to a separation of 1250\,AU and a relative 
movement $\la 15$\,AU over 11\,yr, which results in an upper velocity limit of 6\,kms$^{-1}$ between AB and C. This 
does not rule out the possibility that C has been much closer to AB in the past: With a velocity of 6\,kms$^{-1}$ it 
would have taken only $10^3$\,yr for C to move to its current position from a starting point close to AB.
It is thus conceivable that C has been ejected by a close dynamical encounter with AB in the early evolution 
of the system.

\section{Morphology of IRAS04325}
\label{s4}

\subsection{Structure within IRAS04325}
\label{s41}

The IRAS source is located within the small dark cloud L1535, a part of the B18 cloud complex in Taurus.
The few field stars visible in both H- and K-band have $H-K$ colours of 2-3\,mag, corresponding to extinctions 
of 25-35\,mag, which indicates substantial amounts of gas and dust along the line-of-sight. 

\begin{figure*}
\includegraphics[width=7.5cm,height=8.5cm,angle=-90]{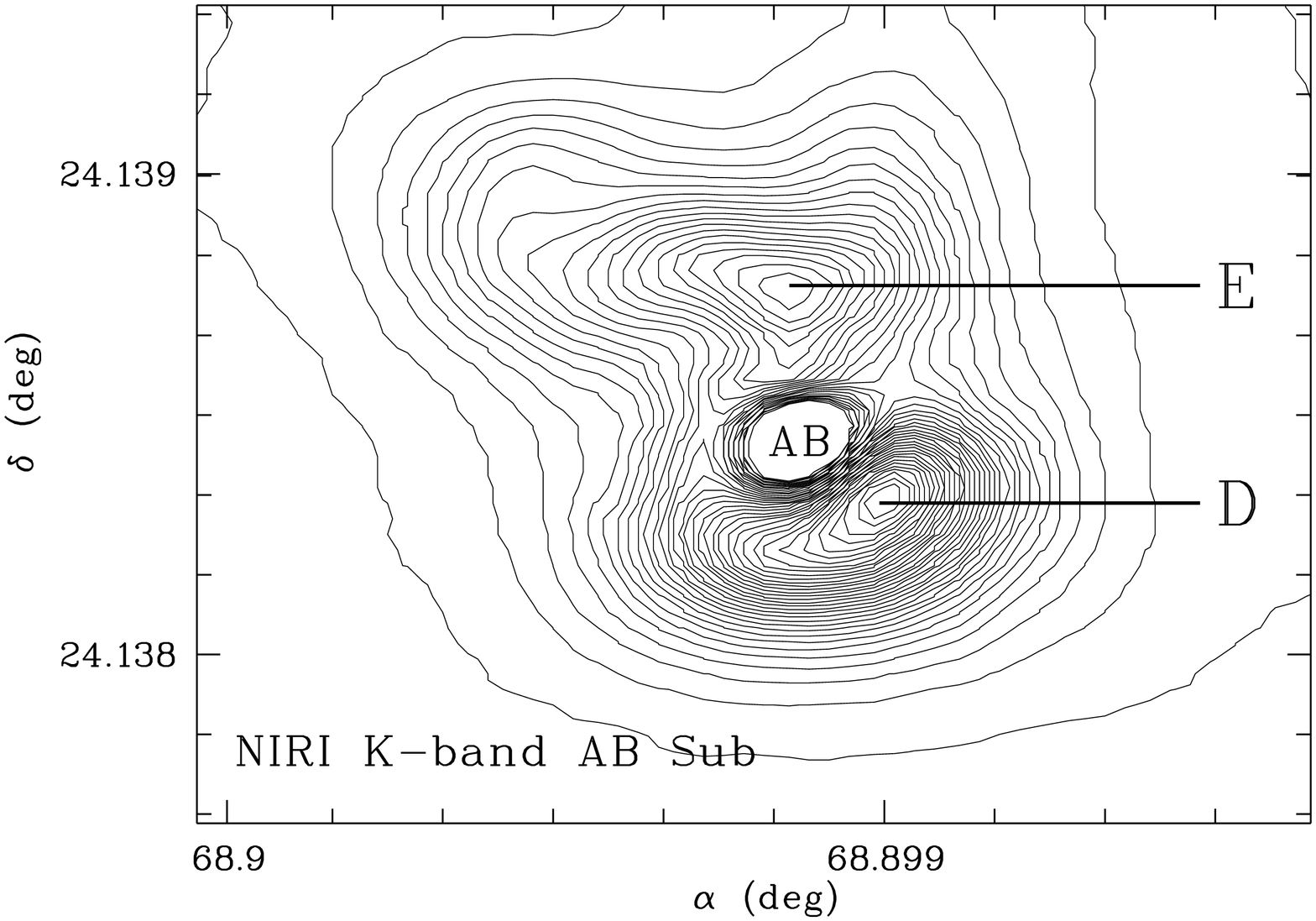} \hfill
\includegraphics[width=7.5cm,height=8.5cm,angle=-90]{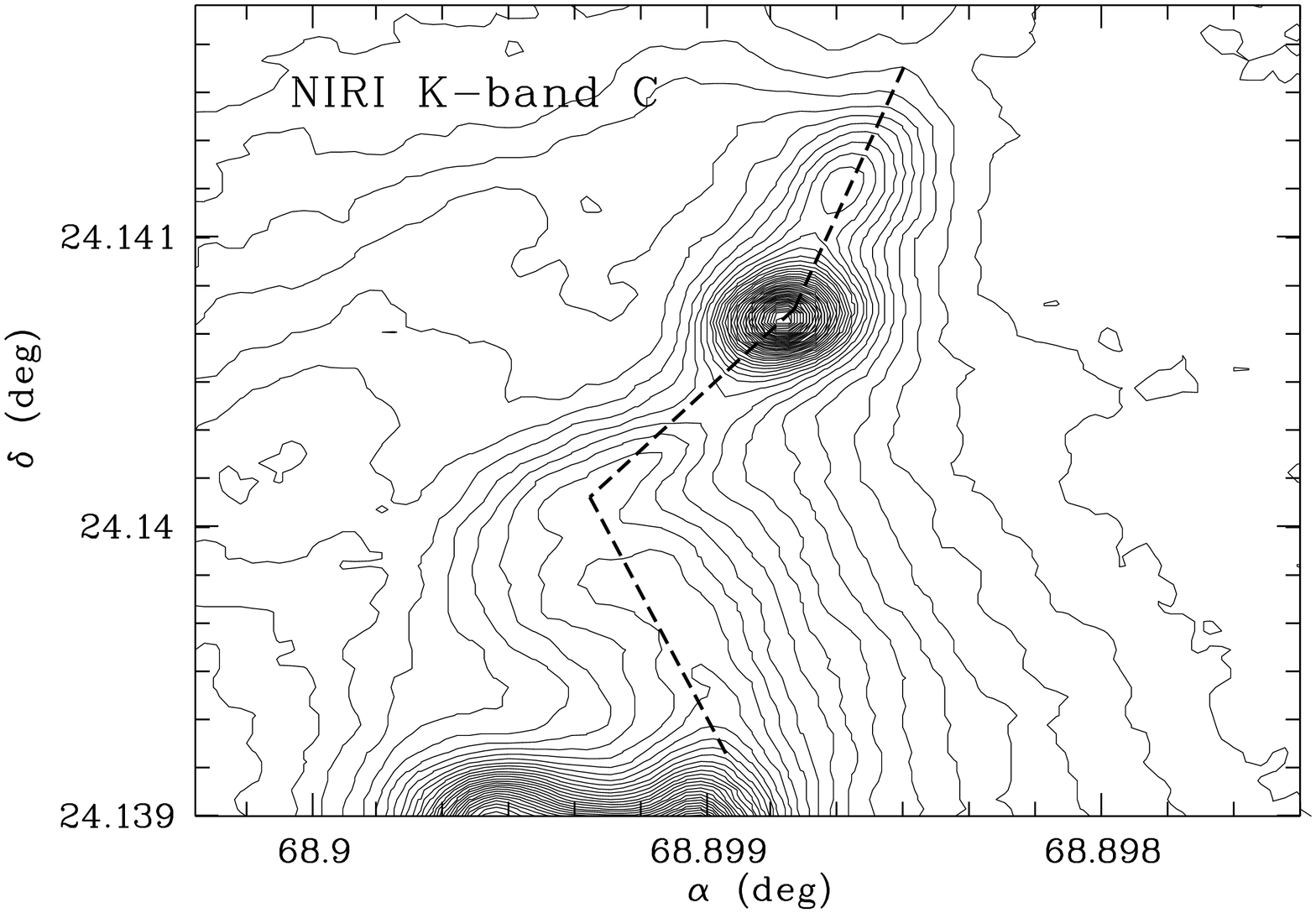} \hfill
\caption{Contour maps of the K-band image from Gemini/NIRI around the source IRAS04325. One small 
tickmark corresponds to 0\farcs6 (5 pixel). The left panel shows a small area around source AB after 
subtracting one point source centered at the peak of the emission. The total image size is $6" \times 6"$. 
The two extended structures D and E to the south and north of the point source are clearly visible. The 
right panel shows the northern part of the source around object C; image size $10"\times 10"$. The 
dashed lines mark the light streaks connecting AB and C and extending further to the north. Note 
that the dark lane from the disk around C, as seen in the HST image, is not resolved here.\label{f2}}
\end{figure*}

The near- and mid-infrared images of IRAS04325 show a bright elongated emission nebula \citep{1994ApJS...94..615H,1999AJ....118.1784H,2007AJ....133.1528C}. In our images in the JHK 
bands the dimensions of this nebulosity are about $10" \times 20"$, corresponding to $1500 \times 3000$\,AU,
with a position angle (PA) of $\sim 15-20$\,deg. The 
nebula is divided in two lobes. While the northern lobe is brighter, the southern one is more extended 
with two long light streaks visible up to 15" to the south of the nebulosity center. The structure is best 
explained as light from object AB scattered by dust at the edges of an outflow cavity; indeed, the orientation 
fits with various outflow related features (Sect. \ref{s42}). 

Apart from the two compact sources, the multi-wavelength images reveal a complex substructure in
IRAS04325, see the contour plot in Fig. \ref{f2}, where we have subtracted a point source at the position
of AB (see Sect. \ref{s34}). We note that the PSF subtraction affects only the area within 0\farcs5 radius
of AB and does not alter the structure seen in the figure. There are two extended features in close 
proximity to AB (Fig. \ref{f2}, left panel). The southern feature, called D in the 
following, has a peak 0\farcs7 SW of AB and appears in the K-band as a half-ring broadly covering the 
south part of the point source. This is best explained by scattered light from an envelope around 
the central source, with a radius of about 100\,AU. The position and structure of this feature 
agrees well with extended CO emission seen in the submm regime at radial velocities of 6.5-8\,kms$^{-1}$, 
indicating the presence of cold gas. In the CO map the maximum intensity is seen at 6.9\,kms$^{-1}$ 
with a peak at 0\farcs8 south of AB (Fig. \ref{f5}). 

The northern feature, called E in the following, appears in the K-band as an open cone with an apex 
located 1\farcs2 north to AB, corresponding to 180\,AU (see Fig. \ref{f0} and \ref{f2}). Two curved light 
streaks extend this structure 
towards the east and west. The same extensions are seen in the submm continuum image (Fig. \ref{f9}). 
The position of feature E also coincides with a CO lobe seen in a velocity regime between 2 and 
5\,kms$^{-1}$ (see Fig. \ref{f5}). It is most pronounced at 4-4.5\,kms$^{-1}$, with a peak position 
at about 1\farcs0 north of AB. Feature E is orientated at a position angle of $\sim 15$\,deg with 
respect to AB, coinciding with the orientation of the emission nebula. The morphology and orientation fit 
well with an interpretation as part of the outflow cavity.

Our own K-band image and the one from HST \citep{1999AJ....118.1784H} show a faint S-shaped structure connecting 
the point sources AB and C and extending further to the north (Fig. \ref{f0} and \ref{f2}, right panel). This is 
most likely scattered light from dust features. It is possible that this 'bridge' between AB and C is coincidental; 
we see the extension of the outflow-related feature E (see above) and light streaks originating from source C in 
superposition in the plane of the sky. \citet{1999AJ....118.1784H} argued that the two faint features 
near C are scattered light from the cavity edges of an outflow from object C. Their direction is 
60-90\,deg offset from the orientation of the disk seen in the HST images, which fits into the 
outflow interpretation. However, in our CO observations there is no evidence for outflow lobes at the
same positions (see below).

Therefore we prefer a scenario in which these structures belong to a physical 'bridge' between AB and C. 
Such a 'bridge' could be indicative of gravitational interaction 
in the early evolution of the system, but the long-term stability of the outflows (Sect. \ref{s42}) argues 
against this interpretation. Instead we propose to explain the structure as dust features from interacting 
accretion streams (as speculated by \citet{1999AJ....118.1784H}). The S-shaped morphology indeed resembles 
the density maps in accreting binaries from the SPH simulations by \citet{1997MNRAS.285...33B}.

Two CO lobes are detected in close proximity to source C in a narrow velocity range of 6.5-7.5\,kms$^{-1}$
(see Fig. \ref{f5}). The strongest one is almost exactly coinciding with the position of the
compact source seen in the continuum image: the offset between peak of the CO intensity and peak in
the continuum is $<1\farcs0$. The PA for the elongation of this CO lobe is roughly 300\,deg. The 
second CO blob is located roughly 5" away at PA $\sim 290$\,deg with respect to C. A third CO detection
is seen in the velocity range of 4.4-4.8\,kms$^{-1}$ 5" west of object C at PA of $\sim 270$\,deg. The
orientation of these three CO features is roughly perpendicular to the alignment of the disk (PA 30\,deg).
Thus, they could be associated with an outflow from source C.

\subsection{Structure outside IRAS04325}
\label{s42}

\begin{figure}
\includegraphics[width=4.1cm,angle=0]{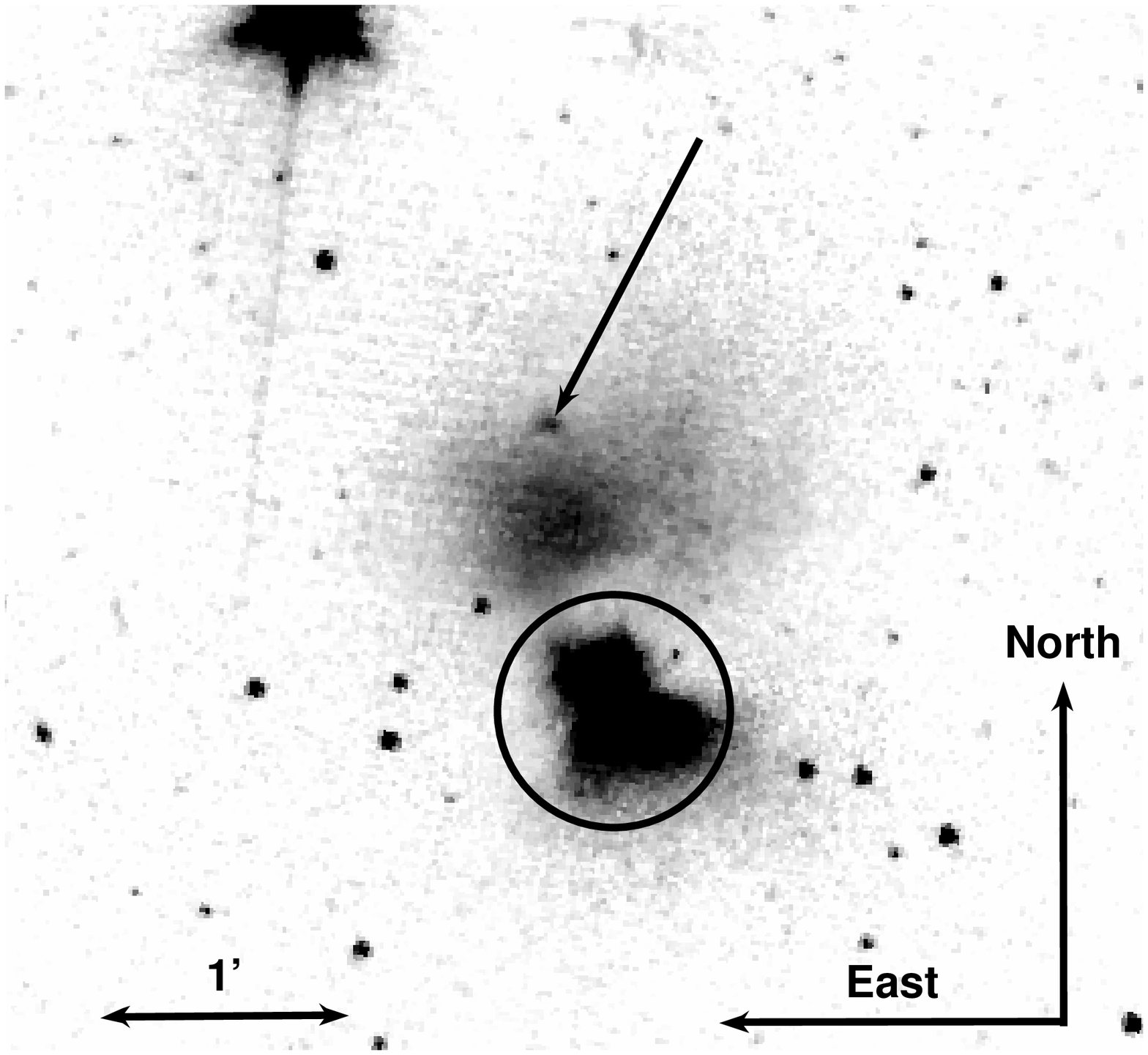} \hfill
\includegraphics[width=4.1cm,angle=0]{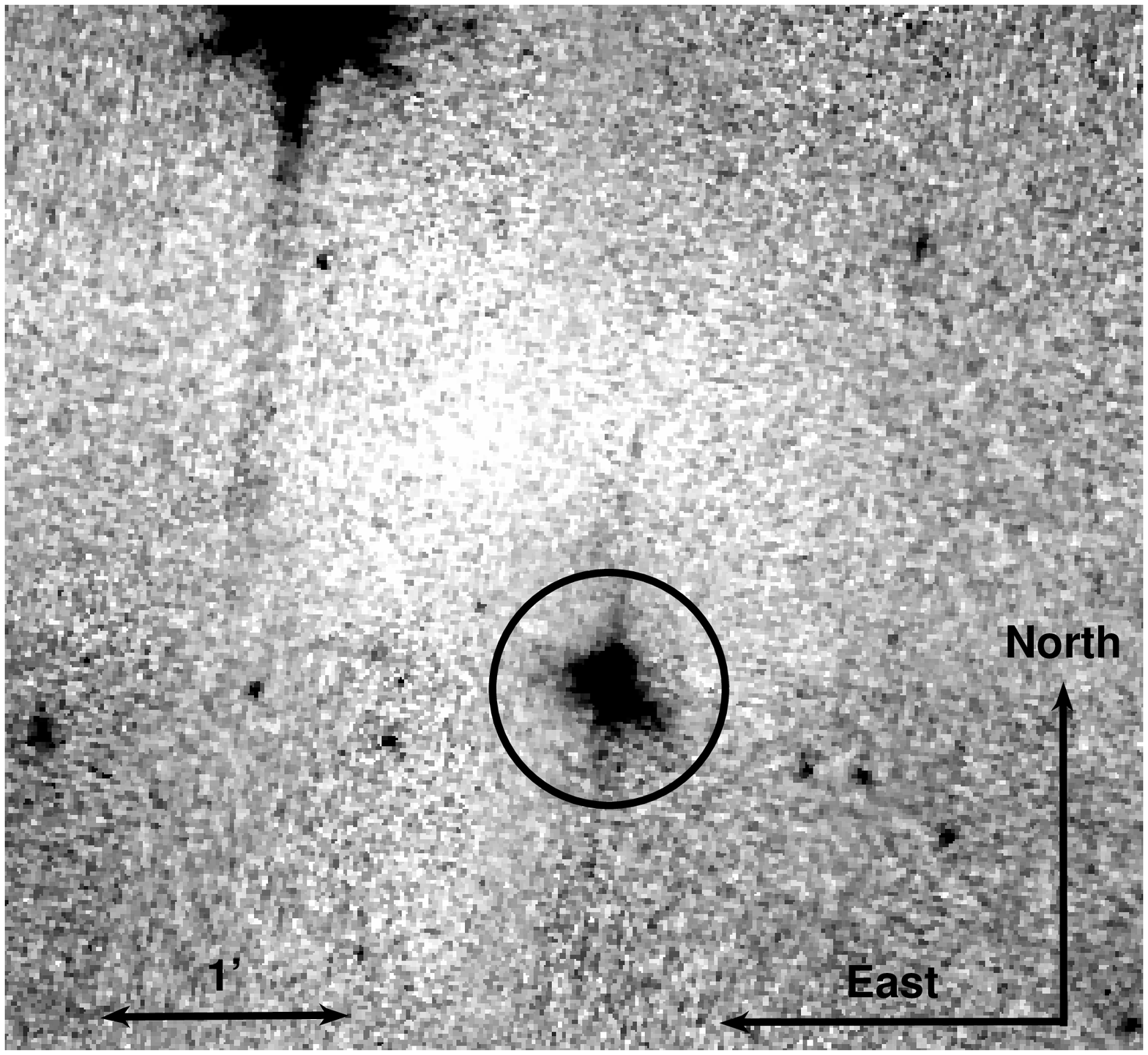} \\
\caption{Spitzer IRAC images centered on the nebulosity north of IRAS04325 at 4.5 (left) and 8.0$\,\mu$m (right). 
IRAS04325 is marked with a black circle. Noteworthy are a) the emission knot seen at 4.5$\,\mu$m 
(left panel, marked with arrow) at the northern end of the nebula and b) the dark feature visible at 8$\,\mu$m. 
\label{f3}}
\end{figure}

A region of diffuse near-infrared emission is found about 40" north and 15" east of IRAS04325AB, corresponding to a
PA of 20\,deg (see Fig. \ref{f3}). Thus, this area is located in the extension of the axis defined by the 
elongated emission nebula. The feature has been mentioned before in the literature, for example by 
\citet{2002AJ....123.3370P}, and is most pronounced at 2 to 5$\,\mu$m. In our K-band image the feature 
has a size of $20"\times 20"$. The proximity to IRAS04325 and the location along the outflow axis lead 
us to believe that the feature is physically associated with IRAS04325. Under this assumption the physical 
distance from IRAS04325 would be $\sim 6000$\,AU. This patch of emission could be scattered light at the 
northern end of the outflow cavity. 

In the IRAC bands at 3.6 and 4.5$\,\mu$m a 10-15" wide, dark band is visible between the IRAS04325 nebula 
and the emission region further north (Fig. \ref{f3}, left panel), possibly a foreground cloud blocking the view
to the source.

At 8$\,\mu$m the bright diffuse patch in the north of IRAS04325 is not visible, instead the image shows a 
roughly circular, $1.5' \times 1.5'$ absorption feature centered 70" north and 30" east of IRAS04325 
(PA 23\,deg), i.e. just north of the bright patch (Fig. \ref{f3}, right panel). This dark feature has 
fainter extensions to the east and south, overlapping with IRAS04325. Its center is located along the 
outflow axis of IRAS04325AB. The position agrees well with the core L1535 N-SMM seen in the submm maps 
of IRAS04325 presented by \citet{2000ApJ...534..880H}, indicating that this feature represents an overdensity 
of dust. It is possible that the northern lobe of the outflow is blocked by this core, explaining the lack 
of large-scale outflow features further to the NE (see below). 

The dark feature is only visible at 8$\,\mu$m, and not at any other wavelength in the mid-infrared. 
In the 8$\,\mu$m image it is the only such feature within a radius of one degree. This may be
related to the presence of strong PAH emission bands in the other IRAC bands \citep[e.g.][]{2002A&A...390.1089P}.
It is conceivable that we see a region of strong absorption projected against the bright PAH background, as
they are often found in the Galactic plane \citep[e.g.][]{1998ApJ...494L.199E}. In fact, the combination
of submm emission and 8$\,\mu$m absorption has been observed for a number of cores in infrared dark clouds 
\citep{2009A&A...499..149V}. 

IRAS04325+2402 has long been known to drive molecular outflows 
\citep{1987ApJ...321..370H,1989ApJ...340..472T,1992ApJ...400..260M}. Specifically, a large body of evidence is 
available for an outflow driven by source AB at position angle of 10-20\,deg. Features associated with this
outflow are the elongated emission nebula \citep[also seen in HCO$^+$,][]{1998ApJ...502..315H}, the scattered 
light feature E (Sect. \ref{s41}), the HH objects 434-436 about 1\,deg south-west (2.7\,pc) of the IRAS 
source at PA of 17.5-19.5\,deg \citep{2001AJ....121.1551W}, and HH703 30' south of IRAS04325 (1.4\,pc) at
PA of 16\,deg \citep{2003ChJAA...3..458S} 

We searched in the H$_2$ images as well as in the IRAC images for more shock features along the outflow axis. 
In particular the IRAC band 2 (4.5$\,\mu$m) is known to contain a number of bright emission lines of shocked
H$_2$ \citep[see][and references therein]{2009ApJ...695L.120Y}. The group of Herbig-Haro objects 434-436, so far only
detected in the [SII] line and possibly part of a bow shock \citep{2001AJ....121.1551W}, is well-visible in 
H$_2$ and in all four IRAC bands. In the 4.5$\,\mu$m image we find an additional knot about 1.1' north of 
AB (10000\,AU) at PA 14\,deg ($\alpha$ $4^h 35^m 36\fs5$, $\delta$ $+24^{o} 09' 25\farcs5$ (J2000), Fig. \ref{f3}, 
left panel). This feature is clearly extended and not seen at any other wavelength, which argues for an association 
with the outflow. No other obvious feature was found for PA of 15-20\,deg and separation from IRAS04325 up to 
1\,deg in both directions.

There is a second large-scale outflow emanating from IRAS04325+2402, identified by \citet{1987ApJ...321..370H}.
The outflow is seen in integrated CO emission spanning 12', i.e. 0.5\,pc if related to IRAS04325. It is 
described as well-collimated, redshifted, and monopolar. In clear disagreement to the first outflow discussed 
above, this second flow is oriented from the IRAS source towards the NW at position angle around $\sim 310$\,deg,
roughly perpendicular to the disk of source C.

The most likely source for this second outflow is object C. This star clearly has jet activity, proven by the
variety of outflow related emission lines in the near-infrared spectrum. Three CO lobes are seen around 
this object with PA of 270-300\,deg, roughly matching the orientation reported by \citet{1987ApJ...321..370H},
which could belong to the outflow as well. We do not find any additional counterparts of this outflow in the 
H$_2$ and the IRAC images.

\section{SED modeling}
\label{s5}

We aim to reproduce the spectral energy distributions (SED) for objects AB and C with a Monte Carlo radiative 
transfer code. In addition to the data presented here and in \citet{2008ApJ...681L..29S} we make use of the 
Spitzer/IRS spectrum available for object AB \citep{2008ApJS..176..184F}. Only the SL exposure is
used, which is obtained through a 3\farcs6 wide slit oriented roughly in N-S direction and covers the wavelength
range from 5.2 to 14$\,\mu$m. This spectrum matches well with the photomery at 11.2$\,\mu$m. We do not use 
the LL spectrum (obtained through a 10\farcs6 slit) and the IRAC/MIPS photometry, due to their poor spatial
resolution.

Our model code computes thermal radiation and non-spherical scattering from dust in a two-dimensional axisymmetric 
system with central source, flared disk, and larger-scale flattened envelope with a narrow evacuated bipolar 
outflow cavity \citep[e.g.][]{2003ApJ...591.1049W,2003ApJ...598.1079W}. The central 
photospheres are approximated with NextGen model atmospheres \citep{1999ApJ...512..377H,2001ApJ...556..357A} 
with $\log{g} = 4$. The main free parameters for the disk are mass, outer radius, inner radius, accretion rate, 
scale height factor $h_0$ and flaring power $\beta$. The latter two are defined as $h = h_0 (r/R_\star)^\beta$, 
with $h$ the scaleheight of the disk and $r$ the distance from the star. The parameters for the envelope include 
envelope accretion rate, mass, outer radius, and outflow cavity angle. In addition the model fits the ambient 
density and the interstellar extinction between the source and the observer. 

For the envelope we assume dust with an interstellar-like size distribution, using the model by 
\citet{1994ApJ...422..164K}. In the disk we use a dust model including silicates and carbonaceous grains 
with solar abundances. Here the grain size distribution is a power law with an exponential decay for 
particles with sizes above 50$\,\mu$m and a formal maximum grain size of 1\,mm \citep{2002ApJ...564..887W}. 
This dust model thus allows for the presence of larger grains than in the ISM. For more details and references 
about the radiation transfer code see \citet{2006ApJ...645.1498S} and \citet{2006ApJS..167..256R}.
The code has recently been extended to treat the non-thermal reprocessing of photons by PAH molecules and 
very small grains \citep[VSG,][]{2008ApJ...688.1118W}, using opacities, abundances, and re-emission templates from 
\citet{2007ApJ...657..810D}. 

Given the large number of free parameters and the numerous simplifying assumptions (e.g., two-dimensional
disk, axisymmetric system, no variability), the modeling is subject to degeneracies 
\citep[e.g.][]{2001ApJ...547.1077C,2007ApJS..169..328R}. The models presented here are
by-eye fits to the SED and do not represent a unique solution. We do not claim to have explored
the full parameter space. Instead, we use the model to verify if the SEDs can in principle be 
understood in the commonly used framework of disk/envelope models. In addition, this will provide 
an independent test for the plausability of the constraints for the physical parameters of the system.
The parameter best constrained by the modeling are the circumstellar mass, as it depends
only on the submm/mm datapoints. On the other hand, the mass distribution in disk and envelope,
the disk inclination, as well as the extinction are poorly constrained.

\subsection{Object AB}
\label{s51}

\begin{figure}
\includegraphics[width=9.0cm,angle=0]{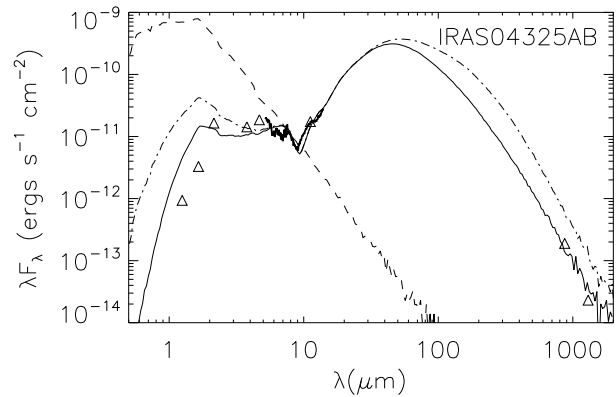} \hfill
\caption{SED fitting for IRAS04325AB. Shown with solid line is a model calculated for a square aperture of 
200\,AU, i.e. comparable to the observations. The dash-dotted line shows the same model for the full scale
of the simulation ($10^5$\,AU). The photospheric spectrum is plotted as dashed line. The observed fluxes
are plotted with triangles, except for the IRS spectrum, which is seen as a thick solid line between 5 and 
14$\,\mu$m. The models assume an external (i.e. outside disk and envelope) extinction of $A_V = 3$\,mag. 
\label{f8}}
\end{figure}

Figure \ref{f8} shows a model for object AB computed without the inclusion of PAH/VSGs. We find that
the two-dimensional model with disk, envelope, and outflow cavity is able to reproduce many of the 
features of the observed SED. One matching SED model is shown in Fig. 
\ref{f8}. This specific model is based on the following parameters, consistent with the observational 
constraints: stellar mass 1\,M$_{\odot}$, stellar radius 2\,R$_{\odot}$, effective temperature 3800\,K, 
disk inclination 78\,deg, disk mass $10^{-3}\,$M$_{\odot}$, disk radius 50\,AU, disk accretion rate
$10^{-8}\,$M$_{\odot}$\,yr$^{-1}$, scale height factor 0.01\,R$_{\star}$, flaring power 1.25, envelope
accretion rate $3\cdot 10^{-6}\,$M$_{\odot}$\,yr$^{-1}$, cavity half opening angle 10\,deg. 
The total circumstellar mass is 0.18\,M$_{\odot}$ for the full scale of the simulation of 
$10^4$\,AU.

These results are consistent with previous attempts to model the source. For example, 
\citet{1997ApJ...485..703W} find a centrifugal disk radius of 50\,AU, an inclination of 72-90\,deg, and 
a total infall rate of $5\cdot 10^{-6}\,$M$_{\odot}$\,yr$^{-1}$ based on near-infrared imaging polarimetry. 
The model presented by \citet{2008ApJS..176..184F} aims to reproduce the full IRS spectrum and broadband 
photometry from the near-infrared to the mm regime and results in a centrifugal disk radius of 100\,AU, 
a cavity semiopening angle of 15\,deg, and a disk inclination angle of 80\,deg.

The largest discrepancies between model and observed SED occur in the near- and mid-IR. This part of the
SED is sensitive to small changes in extinction, stellar parameters, disk inclination and to non-axisymmetric 
geometries \citep{2006ApJ...636..362I}. The images of this object show a complex geometry, i.e. 
3D effects will likely affect the mid-IR emission. 

\begin{figure}
\includegraphics[width=9.0cm,angle=0]{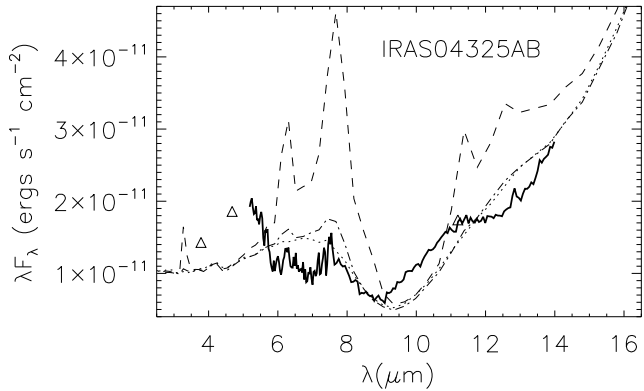} \hfill
\caption{Modeling the IRS spectrum (solid line) for IRAS04325AB. The dotted line is a model without PAH, i.e.
the same as in Fig. \ref{f8}. For the dash-dotted line model, we include PAH in the disk. The
dashed line has PAH in the disk and envelope. Triangles show the photometric datapoints.
All models are calculated using the same parameters as in Fig. \ref{f8}. They assume an external
(i.e. outside disk and envelope) extinction of $A_V = 3$\,mag. 
\label{f8b}}
\end{figure}

The IRS spectrum for this object, shown in detail in Fig. \ref{f8b}, is not well-reproduced by 
our model. The deep absorption trough around 9\,$\mu$m is most likely caused by the amorphous 
silicate absorption band at 9.7$\,\mu$m, as seen in many other protostars \citep{2008ApJS..176..184F}. 
In addition, the spectrum could be affected by the frequently observed ice absorption bands at 6.0 and 
6.8$\,\mu$m \citep[e.g.][]{2004ARA&A..42..119V}.

On the other hand, the peaks at 6.2 and 7.6$\,\mu$m coincide with two of the most prominent features of 
PAH emission spectra \citep{2002A&A...390.1089P}. We made an attempt to model the spectrum by including 
treatment of PAH/VSGs in the radiative transfer simulations. In Fig. \ref{f8b} we show the model with
PAH only in the disk (dash-dotted line) as well as PAH in disk and envelope (dashed line). Qualitatively
the shape of the two peaks is matched by the model.  The absolute flux levels, however, are not adequately 
reproduced by these PAH models. The strength of the peaks is highly sensitive to the location 
of the PAH; in this case, a model with PAH only in the disk provides the best match.

PAH emission is generally found to be rare in disks around young stellar objects, particularly
in late-type objects \citep{2006A&A...459..545G,2006ApJS..165..568F}, with 
detection rates below 10\%. Given the rarity of PAH features and the ubiquity of ice features in 
low-mass protostars, we consider it more likely that ices are predominantly responsible for the 
shape of the mid-infrared spectrum in IRAS04325AB. As our analysis has shown, the presence of PAH 
provides an alternative plausible interpretation for the peaks. To disentangle these effects,
it is necessary to include ice opacities in the models, see for example \citet{2008A&A...486..245C}.

\subsection{Object C}
\label{s52}

\begin{figure}
\includegraphics[width=9.0cm,angle=0]{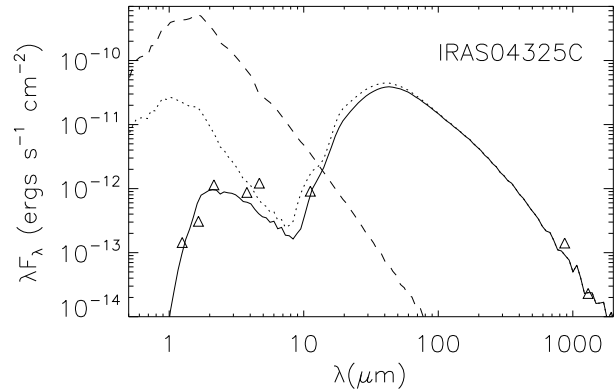} \hfill
\caption{SED fitting for IRAS04325C. Shown is a model with $A_V = 17.5$\,mag that matches
all available independent constraints (effective temperature, disk radius, accretion rate). 
Note that the J-band flux is an upper limit. The dashed line shows the photospheric spectrum, 
the dotted line the intrinsic SED without external extinction.
\label{f7}}
\end{figure}

In Fig. \ref{f7} we show the SED for object C and a model SED that matches most of the datapoints. 
In the modeling we fix the effective temperature of the central object (3400\,K, see
Sect. \ref{s32}), the accretion rate ($10^{-8}$\,M$_{\odot}$\,yr$^{-1}$, see Sect. \ref{s32}), and
the outer disk radius \citep[30\,AU, ][]{1999AJ....118.1784H}. In addition, we require a high
disk inclination ($i>80$\,deg). An edge-on or close to edge-on geometry is necessary to reproduce 
the appearance of the HST image and to explain why photometry and spectrosocpy cannot be simultaneously
reproduced by an extincted photosphere (see Sect. \ref{s33}).

The model shown in Fig. \ref{f7} is calculated for $R_{\star} = 1.8\,$R$_{\odot}$, $i = 85$\,deg, 
$h_0 = 0.01\,$R$_{\star}$, $\beta = 1.25$, i.e. the disk is assumed to be affected by relatively 
strong flaring. The total mass in disk and envelope has to be between 0.001 and 0.01\,M$_{\odot}$
to reproduce the submm/mm datapoints (assuming the canonical gas-to-dust ratio of 100). The 
model predicts most of the
circumstellar mass to be in the disk (0.004\,M$_{\odot}$), with only a thin envelope 
($6 \cdot 10^{-5}\,$M$_{\odot}$), but other configurations are conceivable. The envelope has
an outer radius of 300\,AU, an accretion rate of $10^{-7}\,$M$_{\odot}$\,yr$^{-1}$, and an outflow
cavity with half-opening angle of 15\,deg. 

The model assumes interstellar extinction of $A_V = 17.5$\,mag. This does not include the extinction
caused by disk and envelope, which is extremely high for this object due to the edge-on disk. The 
value for the extinction agrees well with the estimate we obtain from the combined information
of near-infrared spectroscopy and photometry (Sect. \ref{s33}). It differs significantly from the
extinction used in the SED modeling for object AB ($A_V=3$\,mag, Sect. \ref{s51}), which is implausible
because outside the disk/envelope systems the extinction should be comparable. As noted before,
the constraint for $A_V$ obtained from SED modeling is subject to large uncertainties, due
to the numerous degeneracies in the near-infrared SED.

As seen in Fig. \ref{f7}, the photometric datapoint in the M'-band at 4.7$\,\mu$m is difficult 
to reproduce with these models. This does not depend on the particular choice of disk/envelope
properties; the datapoint is an outlier for all reasonable combinations of parameters. Most models
predict a decreasing SED (in $\lambda F_{\lambda}$) between 3 and 5$\,\mu$m, whereas the datapoint
at 4.7$\,\mu$m is 30\% higher than the one at 3.6$\,\mu$m. One possible explanation is additional
flux generated by the outflow, which is not part of the model. The M'-band contains a number of strong $H_2$ 
emission lines typical for shocks in protostellar outflows \citep{2009ApJ...695L.120Y}, in particular 
the 0-0 S(9) transition at 4.69$\,\mu$m. Shocked $H_2$ emission, as seen in the near-infrared for
this object (Sect. \ref{s32}), could contribute significantly to the M'-band flux.

\section{Summary and conclusions}
\label{s6}

We present a multi-wavelength study for the embedded protostellar system IRAS04325+2403. IRAS04325 is 
one of very few known multiple YSOs with at least one resolved disk and two protostellar jets. 
New imaging with subarcsecond resolution has been obtained in the infrared covering the wavelength 
range 1-12$\,\mu$m and in the sub-millimetre at 870$\,\mu$m. This is combined with low-resolution near-infrared 
spectroscopy for the two compact sources in the system and with literature data. In the following we summarize the 
constraints on the properties of IRAS04325:
\begin{enumerate}
\item{We show that the both components in the system are low-mass stars. Object C is an M dwarf, 
probably with an effective temperature around 3400\,K, while AB is likely of earlier type and 
more massive. Object C is unlikely to be a brown dwarf, as speculated previously.}
\item{Despite the wealth of available data, there are enormous difficulties in estimating the 
fundamental parameters of the central sources. In particular, the presence of veiling, disks, and 
strong line-of-sight extinction puts limits on our ability to reliably derive effective temperatures 
from spectra. The poorly constrained age hampers the conversion to masses.}
\item{The near-infrared emission line spectrum for object C provides strong evidence for ongoing
accretion and jet activity.}
\item{We argue that the previously seen double structure of object AB is not due to the presence
of a close stellar companion. Instead it is best explained by an absorption lane in front of the 
object, most likely caused by a disk.}
\item{Comparing our astrometry with the positions reported by \citet{1999AJ....118.1784H}, we put
a limit of $<9$\,mas\,yr$^{-1}$ on the relative proper motion between AB and C. This confirms common
membership in the Taurus star forming region.}
\item{The morphology and SED for object AB is consistent with the presence of a small disk 
(radii $\sim 50$\,AU radius) and a larger envelope ($\la 100$\,AU radius).}
\item{Object C has been found to have an edge-on disk with radius 30\,AU by \citet{1999AJ....118.1784H}.
Our observations confirm that the object is mostly seen in scattered light due to the high inclination.
Based on submm imaging and SED analysis we identify an elongated envelope or outer disk with radius of 
$\la 300$\,AU around this object.}
\item{The mass of circumstellar material in the system is well-constrained by the submm/mm datapoints. 
While the disk masses are quite small (0.001-0.01\,M$_{\odot}$), the total mass within 10000\,AU is
in the range of 0.2\,M$_{\odot}$.}
\item{Many of the features in the SED and the multi-wavelenght images can be understood in the framework 
of commonly used two-dimensional disk and envelope models. The remaining inconsistencies may well be 
caused by the 3D structure of the system, which is visible in the images but not accounted for in the 
models. The mid-infrared spectrum for source AB shows spectral features that cannot be matched with a 
standard model using a single population of dust grains; instead they require the inclusion of silicates, 
ices, and possibly PAHs.}
\item{We identify several new features associated with a parsec-scale outflow driven by object AB with 
position angle of 10-20\,deg. Object C likely has its own large-scale outflow at position angle of 
270-310\,deg.} 
\item{Various pieces of evidence indicate an early evolutionary state for IRAS04325. The 
star-disk systems are embedded in a core with a large line-of-sight extinction of $A_V \sim 30$\,mag. 
The object is undetected in the Taurus XMM X-ray survey \citep{2007A&A...468..353G}, which is unusual 
for Class II objects (15\% undetected), but common for protostars (60\%). The bolometric luminosity 
of the system, as determined by fitting the unresolved SED \citep{2005ApJS..156..169F}, 
is around 1.0$\,L_{\odot}$. According to the pre-main sequence tracks by \citet[][update 1998]{1994ApJS...90..467D}, 
two objects with effective temperatures around 3400\,K would have a total luminosity of 1.3, 1.0, 
0.9$\,L_{\odot}$ at $10^5$, $2\times 10^5$, and $5\times 10^5$\,yr. Thus, an age of a few times $10^5$\,yr 
seems reasonable.}
\item{The position angles of disks, envelopes, and outflows put useful limits on the geometry of the
system. For the AB disk/envelope, C disk/envelope, and binary orbit the position angles are 
approximately 90, 30, and 350\,deg. Thus, the disk/envelope systems in IRAS04325 are strongly misaligned 
with respect to each other (60\,deg) as well as with respect to the orbital plane of the binary (80 and 
40\,deg). The disks are neither aligned nor coplanar.}
\end{enumerate}

The last finding, first stated by \citet{1999AJ....118.1784H} and bolstered here by
more evidence, warrants some further comments. From polarimetry studies it is known that the dominant 
majority (80-90\%) of young binaries has disk orientations within 30\,deg of the orbital plane 
\citep{2004ApJ...600..789J,2006A&A...446..201M}. There are exceptions, found by imaging the jets
\citep[e.g.][]{1994ApJ...437L..55D,2006ApJ...653..425L} or disks \citep{2008ApJ...683..267K}, but 
many of them seem to be higher-order systems where one of the components is again a close binary 
\citep[see the review by][]{2007prpl.conf..395M}. This could be the case for 
IRAS04325 as well, although we argue against the presence of an additional component.

Misalignments can be either primordial or arise during the early evolution, due to dynamical interactions 
in a multiple system. In the case of IRAS04325 the presence of parsec-scale outflows with constant directions 
to within a few degrees seems to exclude strong alignment changes in the recent history. Typical 
dynamical timescales for parsec-scale jets are $10^4$\,yr \citep[e.g.][]{1997AJ....114..280E}. For
the NS jet driven by object AB we estimate a lifetime of $10^4 - 10^5$\,yr, assuming
typical velocities on the order of a few 100\,kms$^{-1}$. The dynamical timescale for the 
NW jet has been determined to be $2\times 10^5$yr \citep{1987ApJ...321..370H}. Thus the system has 
been stable over a substantial fraction of its early history. This is confirmed by the absence of
strong relative motion between the two components, as it might be expected after recent gravitational 
encounters. 

In general the misalignments between disks and orbits in binary systems will decay with time; the 
timescale for this readjustment is expected to be in the order of 20 orbital periods or $10^6$\,yr 
for IRAS04325 \citep{2000MNRAS.317..773B,2004ApJ...600..789J}, which might be longer than its lifetime. For 
higher-order systems the timescales for realignment are much longer, because the various orbital 
planes have to be realigned, not only the disks. In addition, the disk radii in IRAS04325 are 
significantly smaller than the orbital separation ($<50$\,AU vs. 1250\,AU). In such cases it is 
possible that the tilt between disks and orbital plane grows with time and an initial misalignment
is enhanced \citep{2000ApJ...538..326L}. Thus, IRAS04325 might be 
a good case for a system where we see primordial misalignment between disks and orbital plane.

For a wide, isolated system like IRAS04325+2402 there are two main possible avenues of formation: 
fragmentation from a centrifugally supported core (rotational fragmentation) or from turbulent 
material (turbulent fragmentation) -- see \citet{2007prpl.conf..133G} for a review. As discussed 
extensively by \citet{2000MNRAS.317..773B}, primordial misalignments are easy to produce in a 
turbulent fragmentation scenario, for example due to spatial variations in the direction 
of angular momentum within the core. On the other hand, rotational fragmentation predicts a strong 
alignment between disks and orbit. The picture presented here for IRAS04325 shows that strong 
misalignments can be present even in small-scale star forming environments and favours 
turbulent fragmentation as the formation scenario.

\section*{Acknowledgments}
We thank Ian Bonnell for instructive discussions related to subjects discussed in this paper 
and Dirk Froebrich for pointing us to the publicly available $H_2$ images for this region. We
received the IRS spectrum from Elise Furlan who also gave helpful comments on an early version 
of the paper. Her help is greatly appreciated. The careful and constructive review by the
anonymous referee helped to improve the paper significantly.
AS would like to acknowledge financial support from the Scottish Universities of 
Physics Alliance SUPA under travel grant APA1-AS110X. ACG acknowledges financial support from the 
Science Foundation Ireland, grant 07/RFP/PHYF790.
 
\newcommand\aj{AJ} 
\newcommand\araa{ARA\&A} 
\newcommand\apj{ApJ} 
\newcommand\apjl{ApJ} 
\newcommand\apjs{ApJS} 
\newcommand\aap{A\&A} 
\newcommand\aapr{A\&A~Rev.} 
\newcommand\aaps{A\&AS} 
\newcommand\mnras{MNRAS} 
\newcommand\pasa{PASA} 
\newcommand\pasp{PASP} 
\newcommand\pasj{PASJ} 
\newcommand\solphys{Sol.~Phys.} 
\newcommand\nat{Nature} 
\newcommand\bain{Bulletin of the Astronomical Institutes of the Netherlands}

\bibliographystyle{mn2e}
\bibliography{aleksbib}

\label{lastpage}

\end{document}